\documentclass[a4paper]{article}
 \usepackage[dvips]{graphicx}
\usepackage[affil-it]{authblk}
\usepackage[cmex10]{amsmath}
\usepackage[tight,footnotesize]{subfigure}
\usepackage[font=footnotesize]{subfig}
\usepackage{stfloats}
\usepackage{url}
\newlength{\figurewidth}
\usepackage{color}
\begin{document}
\title{
Regional properties of global
communication \\
as reflected in aggregated Twitter data}

\author{
Zs\'ofia Kallus, Norbert Barankai, D\'aniel Kondor, L\'aszl\'o Dobos, \\
Tam\'as Hanyecz, J\'anos Sz\"ule,  J\'ozsef St\'eger, Tam\'as Seb\H{o}k,\\
G\'abor Vattay, Istv\'an Csabai}
\affil{E\"otv\"os Lor\'and University, 
Department of Physics of Complex Systems\\
H-1117, P\'azm\'any P\'eter S\'et\'any 1/A, Budapest, Hungary\\
Email: kallus@complex.elte.hu,barankai@caesar.elte.hu}
\maketitle

\begin{abstract}
Twitter is a popular public conversation platform with world-wide audience and diverse forms of connections between users.
In this paper we introduce the concept of aggregated regional Twitter networks in order to characterize communication between geopolitical regions. We present the study of a follower and a mention graph created from an extensive data set collected during the second half of the year of $2012$.
With a k-shell decomposition the global core-periphery structure is revealed and by means of a modified Regional-SIR model we also consider basic information spreading properties.
\end{abstract}

\section{Introduction}
Twitter is a public conversation site where millions of micro-bloggers post their short messages regularly. As part of the Twitter ecosystem many independent software applications are developed to serve the needs of its diverse user base. This accelerated its growth and allowed the service to establish a world-wide, and  ever growing active user-base. On the other hand, the infrastructure built for the easily automated access to the stream of messages combined with additional public user information, has attracted a scientific audience as well \cite{Twitter1,Twitter2}.  Although many interesting questions can be answered based on the messages and the registered relations of users, it is important to note that the complete data set of public sites is not freely available. In this paper we aim to overcome the sparsity of the accessible data stream by means of different aggregation techniques: over time and within geographic boundaries. As a result, instead of inspecting user-level interactions, we aim to analyze the rather robust communication network of the geopolitical units of the world. Such country-level connections can reveal structural properties and a hierarchy of the regions. By adapting a simple information propagation model we can further differentiate the most efficient information sources of the world - as reflected in high volume of individual conversations.

\section{Methods}
\subsection{Twitter data}
We used a data set collected from the freely available Twitter stream during the second half of the year of $2012$ \cite{TwitterDB}.
We refer to Twitter users in this data set allowing public access to their geographical location information as the geo-users, and each one is located to a single fixed position \cite{TwitterDB}. Those that fell into unmapped territories (\emph{e.g.}, oceans) were discarded from further analysis. By means of aggregation of their different forms of connections we create two different communication networks between geopolitical regions: the mention ($\mathcal{M}$) and the follower ($\mathcal{F}$) networks.

For creating the user-level follower graph we used $177,176,790$ links between $3,312,961$ geo-users identified as the most active ones. Given their list, the additional follower relations were collected separately \cite{TwitterDB}. The source of a following link is the followed user while its target is the follower. A network built using the inverse follower relation is also be meaningful, and can be used for showing the direction of interest. To create the user-level mention graph we used $132,436,279$ mention messages between $5,381,565$ geo-users. The source of a mention link is the sender while its target is the mentioned user.

\subsection{Regional Twitter networks}
The aggregation of a user-level network to obtain a regional one is based on a division of the world into distinct geopolitical units, called regions. We opted to use countries, but in the case of the largest ones (\emph{i.e.} Australia, Canada, USA, Brazil, China, Russia and India) their smaller administrative units are used in order to have a more balanced geographic division. After aggregation the regions become the new nodes of the graph with aggregated user counts, and the new edges are defined directed and weighted based on the respective sums of user level links between regions. For each network regions without any links were discarded from further analysis.

In this paper we aim to characterize the region-level communication. For this purpose both $\mathcal{F}$ and $\mathcal{M}$ are considered as static \footnote{This is justified by their slow evolution compared to the timescale of the relevant processes, and allows for an effective use of the minimal set of freely available Twitter data.}. For creating $\mathcal{M}$ the geographic aggregation is preceded by an aggregation over time. This gives robustness against possible sampling fluctuations coming from the freely available Twitter stream.
	
\subsection{Graph properties} 
Both $\mathcal{F}$ and $\mathcal{M}$ are represented by weighted and directed graphs. Their adjacency matrices $F$ and $M$ have non-negative entries and they are not symmetric. The largest entries in a row of $F$ or $M$ are usually localized in the diagonal, so a careful treatment of the regional and inter-regional communication is necessary.\par
To describe inter-regional communication, we introduce the following characteristics. For the region, labeled by $i$, the Total Volume of Incoming Followers (Mentions) is the sum of the off-diagonal entries in the $i$th column of the Follower (Mention) matrix and is denoted by $\mathrm{TVI}_F(i)$ ($\mathrm{TVI}_M(i)$). The Total Volume of Outgoing Followers (Mentions) is the sum of the off-diagonal entries in the $i$th row of the Follower (Mention) matrix and is denoted by $\mathrm{TVO}_F(i)$ ($\mathrm{TVO}_M(i)$). $\mathrm{TVI}_F$ and $\mathrm{TVO}_F$ of a region describe the total amount of channels of passive information flow originated in the communication environment of the region or in the region itself, respectively. $\mathrm{TVI}_M$ and $\mathrm{TVO}_M$ of a region are related to the activity of the region in conversation based active information flow - incoming and outgoing mentions describe how many conversation has been initiated in the communication environment of the region and in the region itself, respectively.\par
For a given region, $\mathrm{TVI}_F$ and $\mathrm{TVO}_F$ are not necessarily the same. If such an asymmetry occurs, the region becomes mainly information receiver or transmitter. The same is possible for the mention measures. Here, the asymmetry between the values of $\mathrm{TVI}_M$ and $\mathrm{TVO}_M$, describes different attitudes of a region with respect to starting conversations on Twitter. To measure asymmetry in the Follower and Mention matrices, we define type I and type II asymmetry parameters. Type I asymmetry parameter is defined for regions with $\mathrm{TVO}_F$ ($\mathrm{TVO}_M$) greater than zero and it is the fraction $\mathrm{TVI}_F/\mathrm{TVO}_F$ ($\mathrm{TVI}_M/\mathrm{TVO}_M$). Type II asymmetry parameter is defined for regions with $\mathrm{TVI}_F$ ($\mathrm{TVI}_M$) greater than zero and it is the fraction $\mathrm{TVO}_F/\mathrm{TVI}_F$ ($\mathrm{TVO}_M/\mathrm{TVI}_M$). \par
For further analysis of inter-regional communication, we define $\hat{F}$ and $\hat{M}$, the symmetric, unweighted Follower and Mention matrices by setting all diagonal entries of $F$ ($M$) to zero and setting an off-diagonal entry to one if and only if at least one following (mention) between the corresponding regions has been recorded by our data mining methods. The matrices $\hat{F}$ and $\hat{M}$ can be recognized as adjacency matrices of unweighted graphs. We used $k$-shell decomposition of these graphs to determine their centrality properties. \par
$k$-shell decomposition is a standard tool in graph theory to detect the core-periphery structure of a graph \cite{kS1},\cite{kS2},\cite{kS3}. A set of nodes $\mathcal{C}_k$ is called $k$-core of the graph $\mathcal{G}$ if the degree of nodes in $\mathcal{C}_k$, with respect to the subgraph of $\mathcal{G}$ induced by $\mathcal{C}_k$, is at least $k$ and $\mathcal{C}_k$ is the largest set of nodes of $\mathcal{G}$ with this property. The $k$-shell $\mathcal{S}_k$ of the graph is defined by $\mathcal{S}_k=\mathcal{V}_k\setminus\mathcal{V}_{k+1}$. Throughout the paper we will say that the nodes of the shell with the highest $k$ belong to the \textit{core} and all other nodes belong to the \textit{periphery}.\par

To relate the regional and the inter-regional communication to each other, we introduce the Normalized Interest Measure (NIM) and the Normalized Activity Measure (NAM) for the $i$th region as 
$$\mathrm{NIM}_X(i):=\frac{\sum_{j\neq i}X_{ji}}{X_{ii}\sum_{j}\hat{X}_{ji}}\quad\mathrm{NAM}_X(i):=\frac{\sum_{j\neq i}X_{ij}}{X_{ii}\sum_{j}\hat{X}_{ij}}$$
if the fractions are meaningful. Here, $X$ is the Following or the Mention matrix. Intuitively, $\mathrm{NIM}_X(i)$ and $\mathrm{NAM}_X(i)$ is the empirical probability to find a following or a mention  between region $i$ and one of its randomly chosen neighbors in the data set, if a following or a mention is found between two users, both of them localized in region $i$, in the same data set.\par
    
\subsection{Regional-$SIR$ dynamics on aggregated networks}

To compare world-wide information spreading potential of regions we adapt one of the simplest propagation models, the $SIR$ model \cite{sir1},\cite{sir2},\cite{sir3}, and modify it to be used on aggregated regional networks. The regional-SIR ($R$-$SIR$) model also assumes that at any given time $t$ an agent can be in one of the following states: susceptible, infected or recovered. The dynamics of single agents is governed by the rate of contact between agents, infection of a contacted susceptible agent and recovery of an infected one. After aggregation of single contact links the new weights are proportional to contact frequencies between pairs of regions on average. By fixing their absolute value a timescale can be defined.

Let $X$ denote either $M$ or $F$ - fixing the total number of regional nodes ($N$) and the regional population counts ($N_k$) accordingly. The state of the $k$th node is characterized by the number of its agents being in each of the three different states: $N_k = S_k + I_k + R_k$. $Y_k$ is used as a compact notation for any of the three variables. We define the normalized edge weights to obtain region-level contact rates per source and per target agent: $x_{kj} := (N_k \cdot N_j)^{-1} \cdot X_{kj}$. 

A few intuitive rules can be translated into a set of deterministic equations governing the time evolution of the regions. We assume that the position of each agent and their total number per region are constant and contact is only possible between two agents in regions connected by an aggregated edge of non-zero weight. The number of contacts per unit time between two regions is assumed to be proportional to the $X_{kj}$ weight of the connecting edge\footnote{Direction of links and regional self-loops are taken into account.} and susceptible agents can only receive information when contacted by infected agents. The number of infected agents decreases from recoveries at a given rate. After division by $N_k$ the relative regional state variables: $y_k(t) := N_k^{-1} \cdot Y_k(t)$ and the new weight matrix: $w_{kj} := x_{kj} \cdot N_j$ will appear in the corresponding equations:
\begin{equation}\label{eq:RSIR}
\begin{aligned}
\frac{ds_k}{dt} &= -\alpha_k \cdot \sum_j{\left( w_{kj} \cdot i_j\right) }\cdot s_k \\
\frac{di_k}{dt} &= +\alpha_k \cdot\sum_j{\left(w_{kj} \cdot i_j\right) }\cdot s_k - \beta_k \cdot i_k \\
\frac{dr_k}{dt} &= + \beta_k \cdot i_k.
\end{aligned}
\end{equation}
$\alpha_k$ and $\beta_k$ scale the regional characteristic time of infection and recovery respectively. These parameters can further account for cultural and/or topic-related local differences, but in this paper we assume homogeneous distribution across the regions: $\beta_k=\beta$ and $\alpha = 1$, $\forall k$. We define the regional total infection at time $t$ as the sum of infections up until $t$ regardless of possible recoveries
. For describing the global state we use average variables calculated from the global sums $\langle y \rangle _{gl}(t) := N^{-1} \sum_{k}{y_k(t)}$. The global infection time of the $k$th region ($\mathrm{GIT}_k$) is the time needed for $\langle i_{tot} \rangle_{gl}(t)$ to reach $99\%$ with $\beta=0$ and only the $k$th $i_{0,k} = 1$ set to non-zero initial value\footnote{Values are rounded to the resolution of the simulations.}.

\section{Results}

\subsection{Properties of Twitter graphs}
The Follower matrix has been built with the usage of approximately $1.77\cdot 10^8$ followings between $473$ regions. $57.17\%$ of the followers addressed a Twitter user from his or her own region. The sparsity of $F$ is approximately $0.49$. The Mention matrix contained approximately $1.32\cdot 10^8$ mentions from $476$ regions. $82.72\%$ of these mentions addressed a Twitter user from his or her own region. The sparsity of $M$ was $0.29$. \par
Both $\mathrm{TVI}_F$ and $\mathrm{TVI}_M$ reached their maximal value at California. The region with the maximal volume of $\mathrm{TVO}_F$ and $\mathrm{TVO}_M$ was United Kingdom. It seems that California is the most active information source - $9.35\%$ of the total number of followings addressed a user located in California and $5.53\%$ of the mentions were used to try to initiate a conversation with a Californian user. On the contrary, United Kingdom is the most effective information gatherer - $6.27\%$ of the followings have the island country as origin and $5.25\%$ of the mentions have been made by a user from United Kingdom.\par

\begin{figure}[!h]
\centering
\includegraphics[trim=0in 0.65in 0.in 0.75in, width=3.5in]{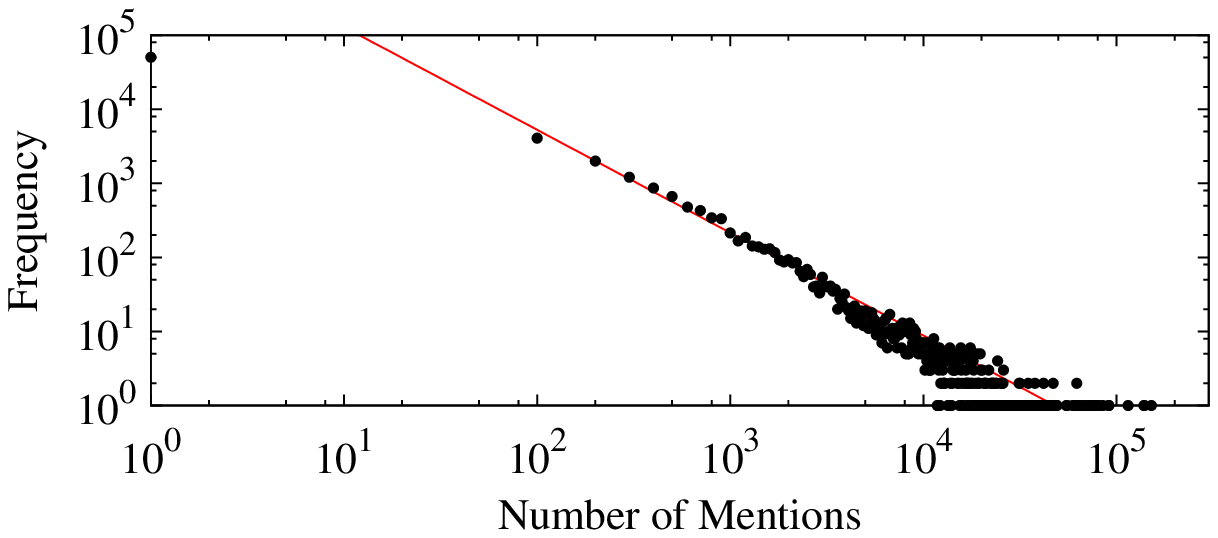}
\includegraphics[trim=0in 0.65in 0.in 0.75in, width=3.5in]{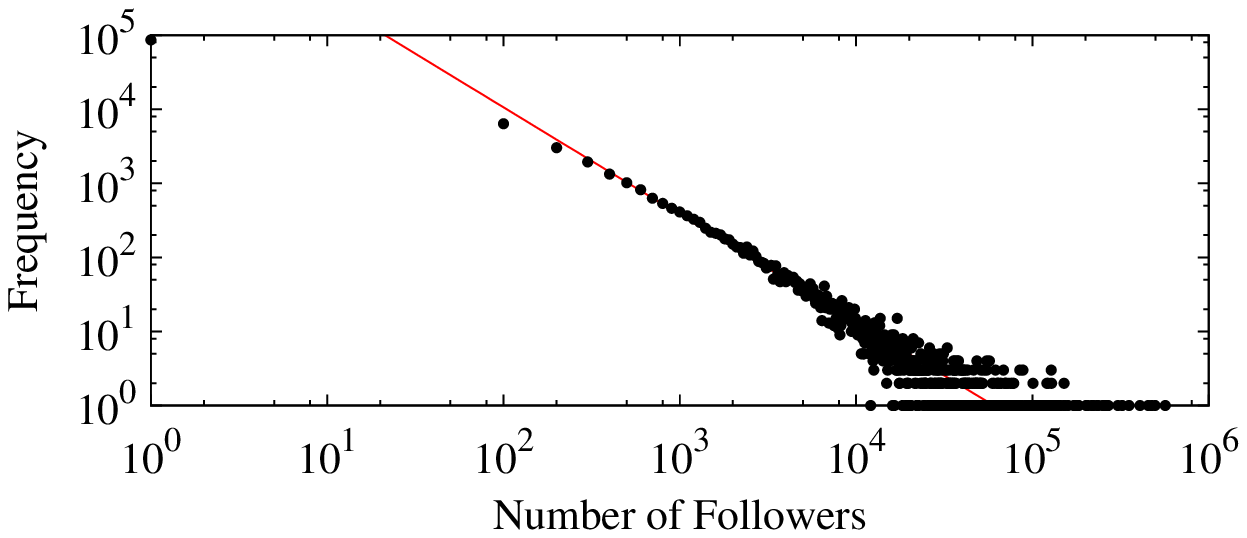}
\caption{Histograms of the empirical distribution of the off-diagonal entries of the Follower (top) and Mention (bottom) matrix.}
\label{OffdiagDistrib}
\end{figure}

It is interesting to ask the properties of the empirical distribution of the non-zero off-diagonal entries of $F$ and $M$. We found that these distributions closely follow an $\sim x^{-\alpha}$ type power law. Cutting away the entries with extremely small and extremely large magnitude, the best fit to the remaining data gave $\alpha_F=1.44\pm 0.02$ in the case of the Follower and $\alpha_M=1.38\pm 0.03$ in the case of the Mention matrix (see Fig.~\ref{OffdiagDistrib}).\par 

\begin{figure}[!h]
\centering
\includegraphics[trim=0in 0.65in 0.in 0.75in, width=3.5in]{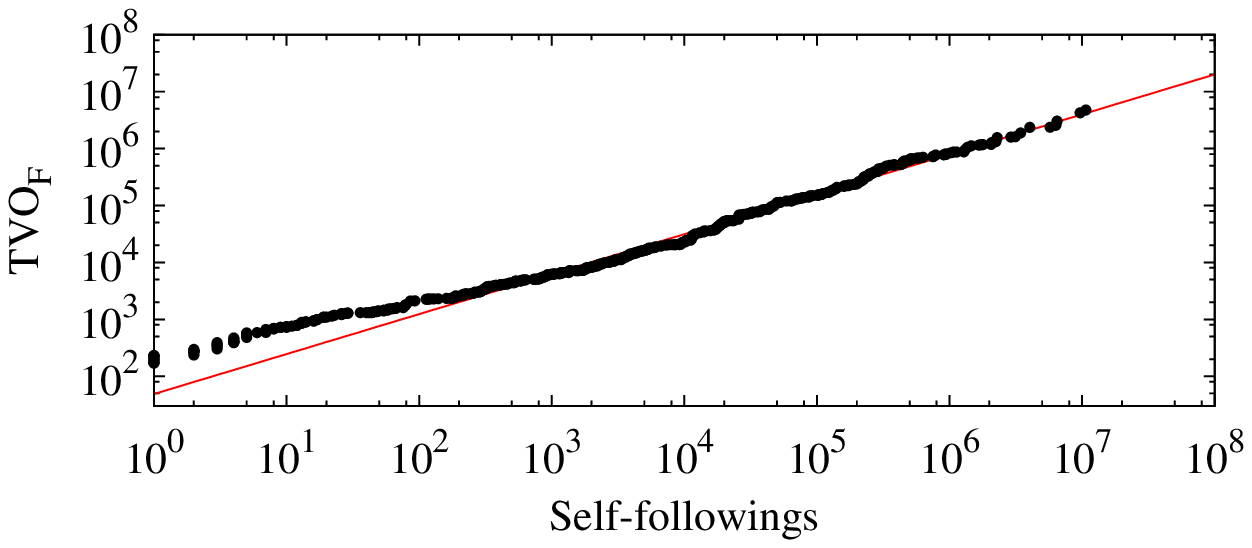}
\includegraphics[trim=0in 0.65in 0.in 0.75in, width=3.5in]{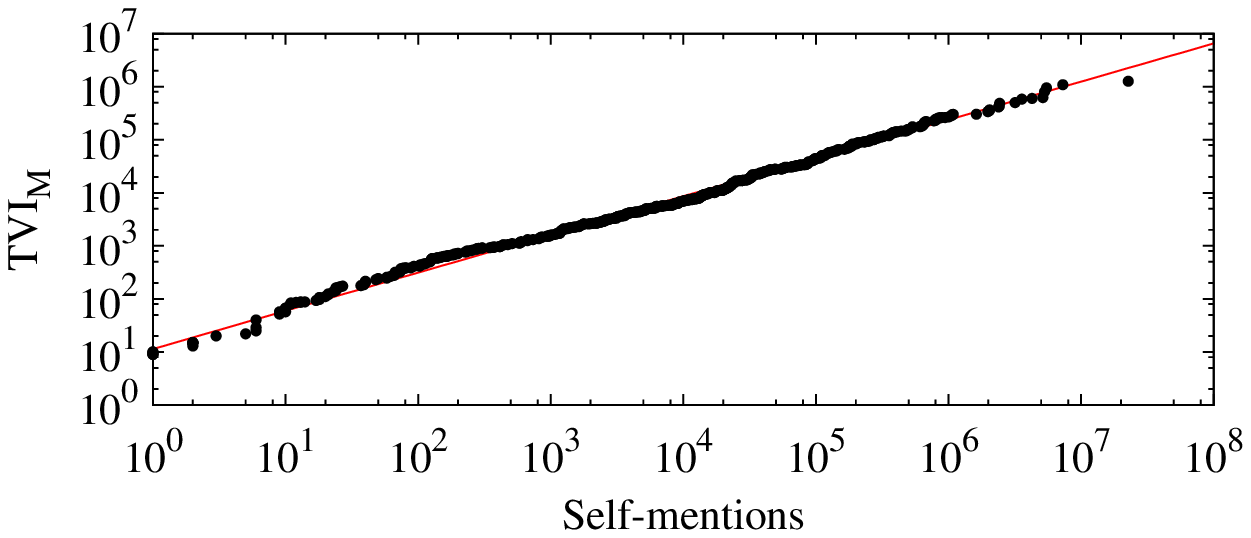}
\caption{Values of $\mathrm{TVI}_F$ and $\mathrm{TVO}_M$ as a function of the number of self-followings and self-mentions.}
\label{SelfVSFollowerMention}
\end{figure}

As it is expected, higher value of self-followers of a region results in higher values of $\mathrm{TVO}_F$ and $\mathrm{TVI}_F$ of the same region. The same is true in the relationship of the self-mentions and $\mathrm{TVO}_M$ and $\mathrm{TVI}_M$. Surprisingly, the dependencies also follow power laws (see Fig.~\ref{SelfVSFollowerMention} in the case of $\mathrm{TVO}_F$ and $\mathrm{TVI_F}$). The exponent in the power law of the best fit to the data using the regions with at least one hundred and at most one million incoming or outgoing followers (mentions) in total is shown in Table \ref{SelfVSMentionFollowFit}. This power law dependence could be the result of observations described in \cite{2005Densification}, but further studies would be needed to decide in reason of the aggregated nature of our data sets.

\begin{table}[!h]
\renewcommand{\arraystretch}{1.3}
\caption{The exponents of the power law dependence of $\mathrm{TVO}_X$ and
         $\mathrm{TVI}_X$ as the function of self-communication.}

\label{SelfVSMentionFollowFit}
\centering
\begin{tabular}{|c||c|c|}
\hline
Type & $F$ & $M$\\
\hline\hline
TVI & $0.710\pm 0.004$ & $0.721\pm 0.003$\\
\hline
TVO & $0.702\pm 0.004$ & $0.830\pm 0.002$\\
\hline
\end{tabular}
\end{table}

\begin{figure}[!h]
\centering
\includegraphics[trim=0in 0.65in 0.in 0.75in, width=3.5in]{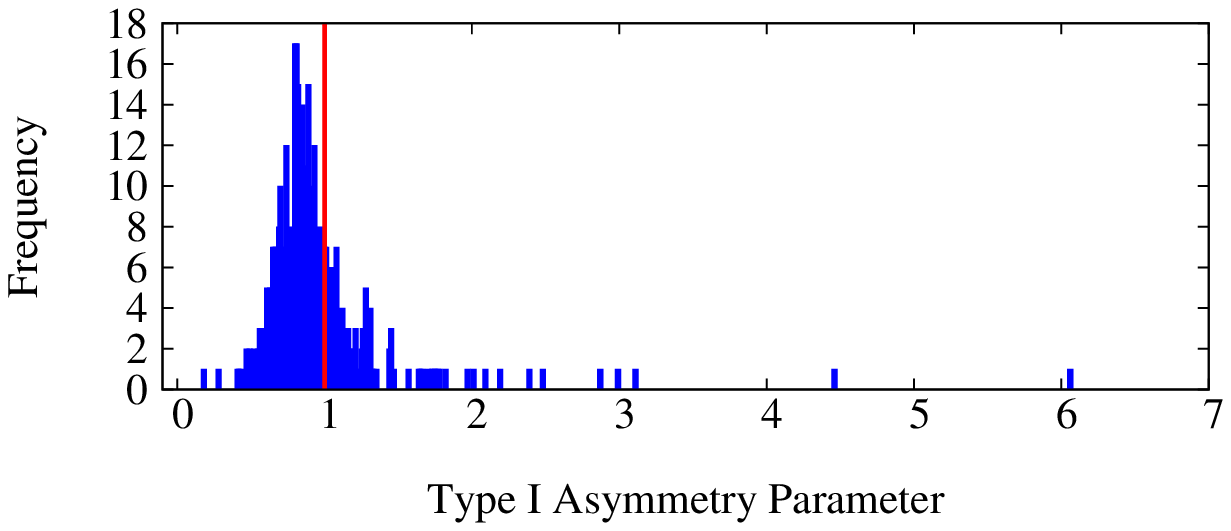}
\includegraphics[trim=0in 0.65in 0.in 0.75in, width=3.5in]{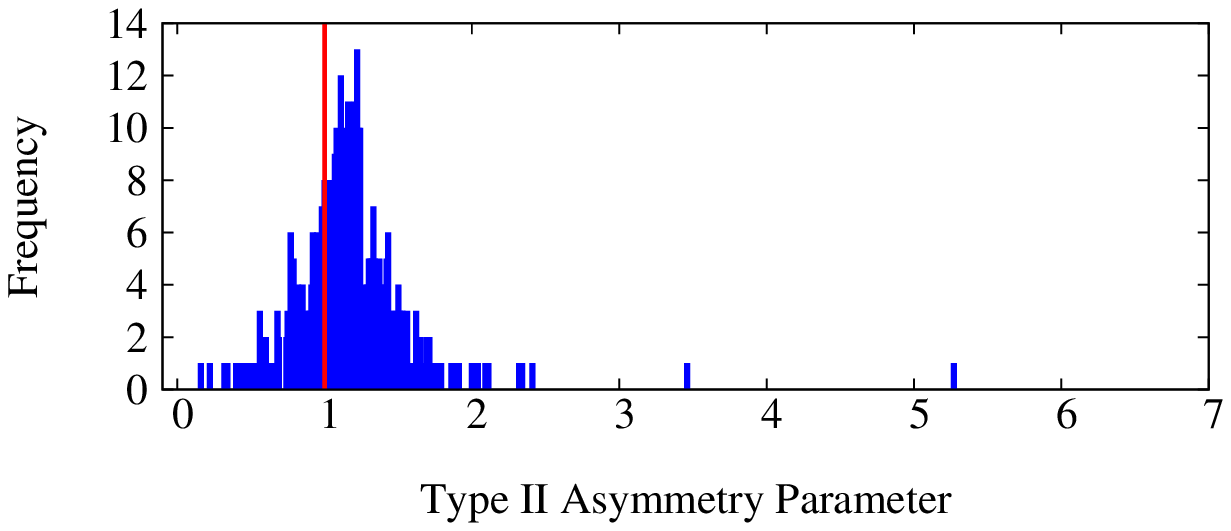}
\caption{Histograms of the empirical distributions of asymmetry parameters in the case of Follower matrix. The $x=1$ curve is marked by a red straight line.}
\label{FAsymmetry}
\end{figure}

\begin{figure}[!h]
\centering
\includegraphics[trim=0in 0.65in 0.in 0.75in, width=3.5in]{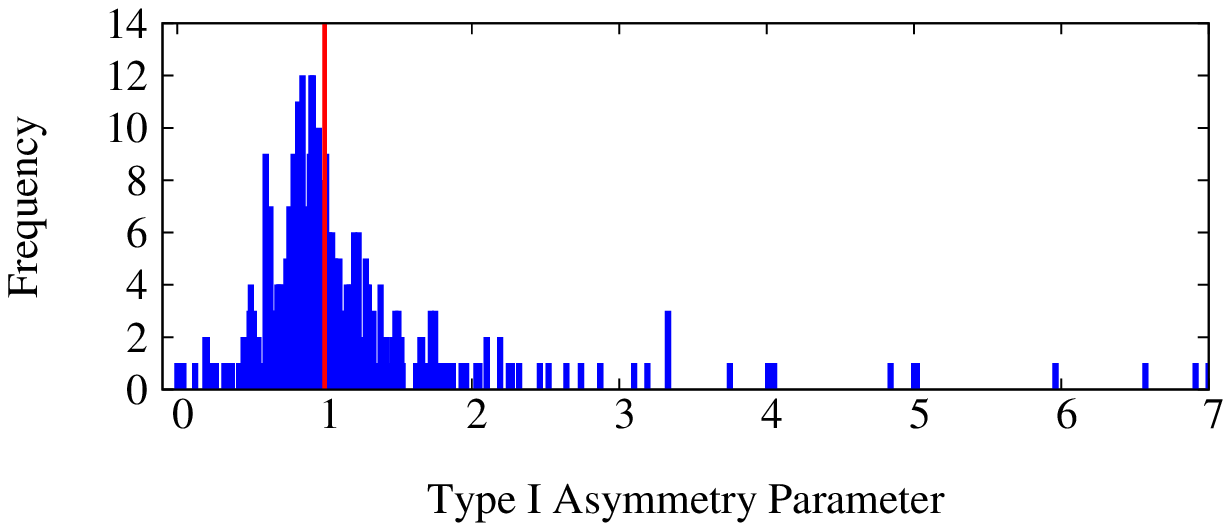}
\includegraphics[trim=0in 0.65in 0.in 0.75in, width=3.5in]{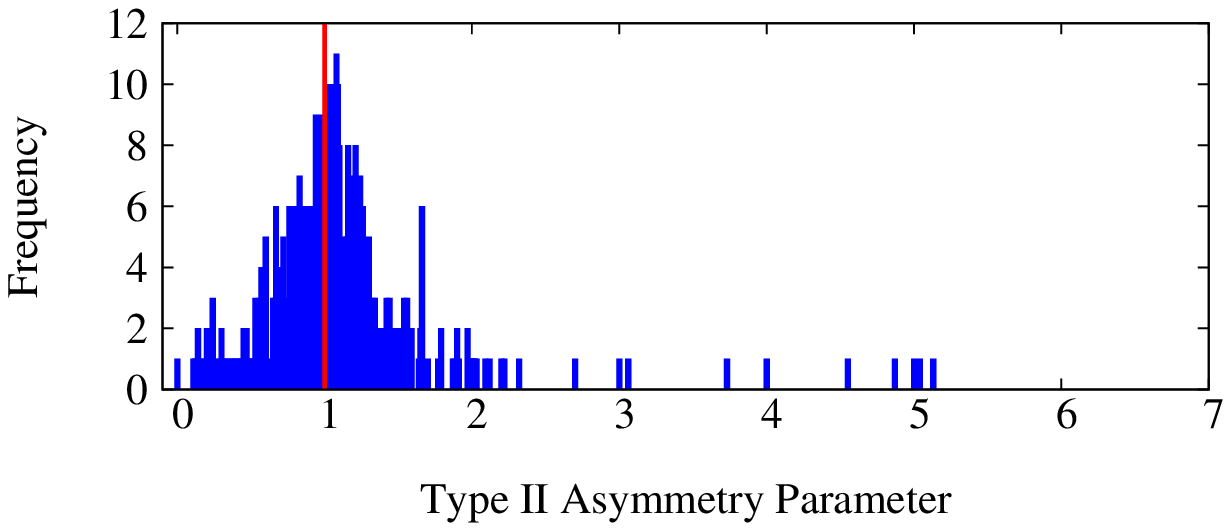}
\caption{Histograms of the empirical distributions of asymmetry parameters in the case of Mention matrix. The $x=1$ curve is marked by a red straight line.}
\label{MAsymmetry}
\end{figure}

We calculated the asymmetry measures for all regions, where they could be defined. We found that most of the regions have type I and type II asymmetry parameters close to one, but there were serious deviations - $14.6\%$ and $18.4\%$ of the regions had type I and type II follower asymmetry parameter greater than one. The same ratios are $19.1\%$ and $20.0\%$ in the case of type I and type II mention asymmetry parameters. The median of type I and type II asymmetry parameter is $0.86$ and $1.03$ in the case of the Follower matrix and $0.96$ $1.03$ in the case of the Mention matrix (see Fig.~\ref{FAsymmetry} and Fig.~\ref{MAsymmetry}). The most frequent type I and type II asymmetry parameters were located around  $0.81$ and $1.22$ in the case of the Follower matrix and $0.86$ and $1.08$ in the case of the Mention matrix.

\begin{figure}[!h]
\centering
\frame{\includegraphics[width=3.5in]{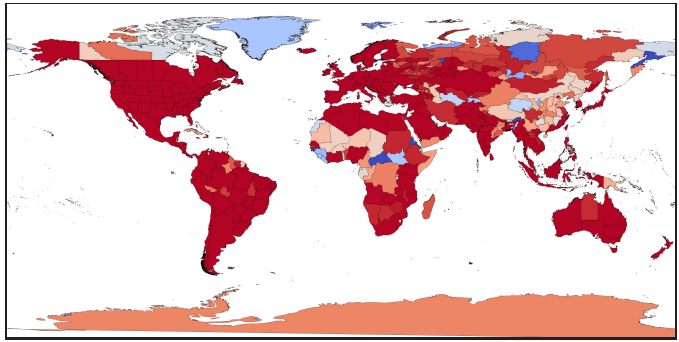}}
\caption{$k$-shells of the symmetric, unweighted Follower matrix on the world map. Deepest blue marks the $k=1$ shell, deepest red is the core.}
\label{kshellworld}
\end{figure}

\begin{figure}[!h]
\centering
\frame{\includegraphics[width=3.5in]{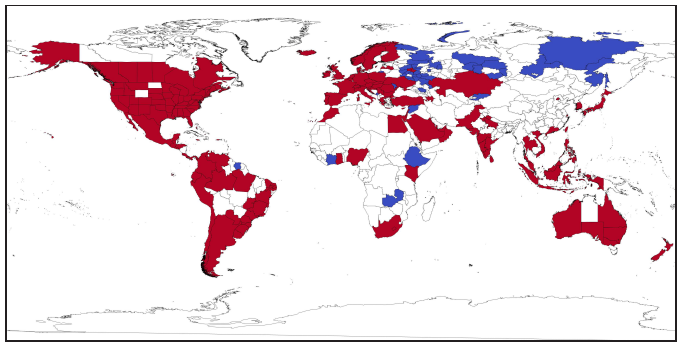}}
\caption{The core (red) and the largest shell in the periphery (blue) of the graph defined by the symmetric, unweighted Mention matrix.}
\label{exceptionalshell}
\end{figure}

The $k$-shell decomposition of $\hat{F}$ and $\hat{M}$ resulted in a massive core and a heavily articulated periphery (see Fig.~\ref{kshellworld}). In the case of $\hat{F}$, the core contains $240$ regions, each of them have at least $199$ neighbors. In the case of $\hat{M}$, the core contains $173$ regions, each of them have at least $135$ neighbors. There is a strong difference between the structure of the periphery of $\hat{F}$ and $\hat{M}$. The periphery of $\hat{F}$ contains $233$ regions and $104$ shells, and the average number of the regions in each periphery shell is low. On the contrary, the periphery of $\hat{M}$ has one large shell and several small shells with only a few regions. The large shell in the periphery contains $45$ regions, each of which have at least $112$ neighbors in the graph of $\hat{M}$ (see Fig.~\ref{exceptionalshell}). 

\begin{figure}[!h]
\centering
\includegraphics[trim=0in 0.65in 0.in 0.75in, width=3.5in]{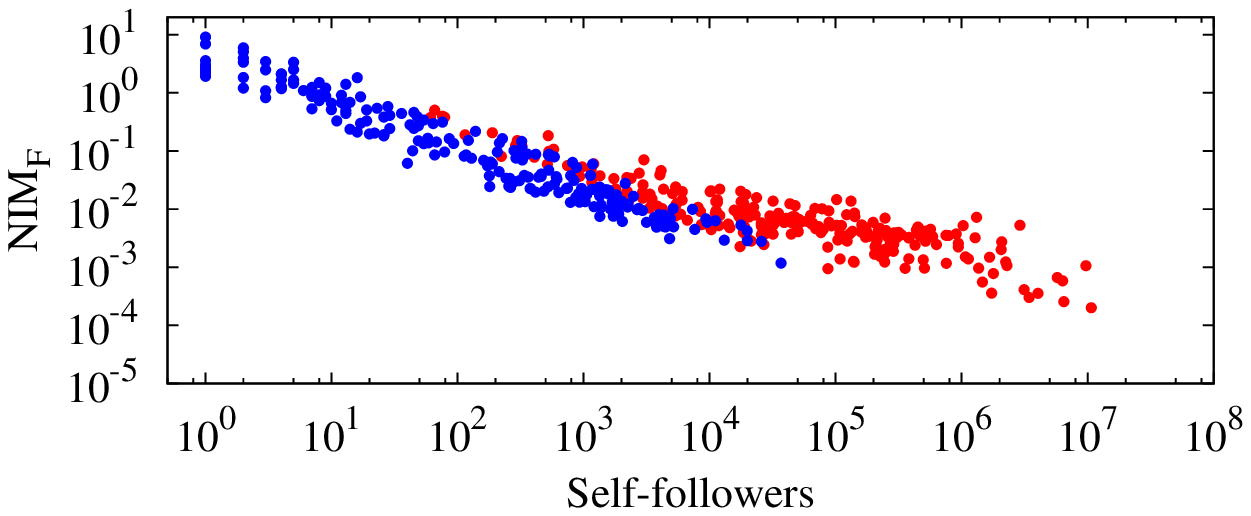}
\includegraphics[trim=0in 0.65in 0.in 0.75in, width=3.5in]{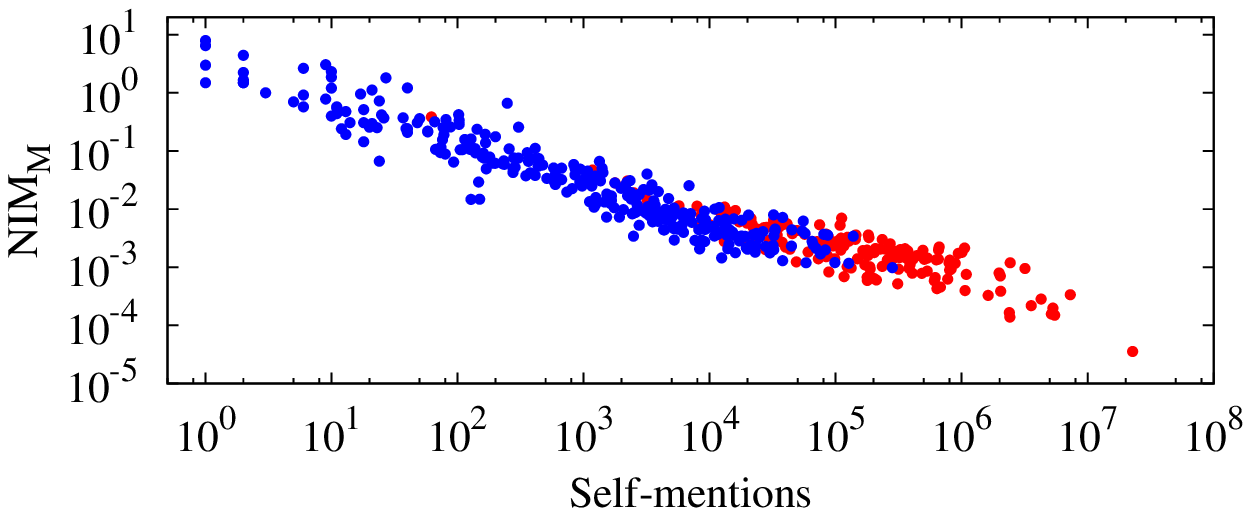}
\caption{Values of NIM in the Following (top) and in the Mention matrix (bottom), as a function of self-followings. Regions of the core and the periphery is marked by red and blue dots. }
\label{NIMMF}
\end{figure}

We end our discussions about the structural properties of the Follower and the Mention graph by the examination of the NIM and NAM parameters. Intuitively, if a region is massively represented on Twitter, its inter-regional activity is high. Our aim with the introduction of NIM and NAM was to compare the trend of the alteration of the intensity of inter-regional communication, if we compare it to the activity of the regional communication. In order to achieve this, we studied NIM and NAM as the function of self-mentions and self-followers. We found that NIM and NAM decays along increasing regional communication. This is a rather surprising result, which means that the growth of the activity in the inter-regional communication with a randomly chosen neighbor of the communication environment of the region increases slower compared to the growth in the number of self followers and self mentions. We found that the rate of the decay is different along the periphery and the core. It seems that both of them follows a near power law, but with our current data, we cannot say surely that the data set comes from the sampling of two different power law or from one convex function which differs from power law (see Fig.~\ref{NIMMF}).

\subsection{Regional SIR dynamics on the Twitter networks}
\label{sec:results_RSIR}
We performed simple $R$-$SIR$ simulations with the primary goal of comparing the global spreading potential of regions. Thus the initial conditions of a single run were set to a single source region being fully infected while the rest of the world was set to susceptible populations. In each case we performed pairs of simulations (differing only in their time interval) using $F$ and $M$ for creating the respective $w$ weight matrices of Eq. \ref{eq:RSIR}.

\subsubsection{Dependence on the recovery rate}
\begin{figure}[!h]
\centering
\includegraphics[width=3.5in]{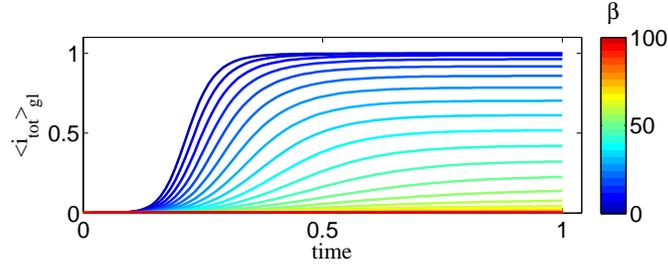}
\caption{Effect of increasing $\beta$ parameter on $\left\langle i_{tot} \right\rangle_{gl}(t)$. $F$ was used with Monaco set as the initial fully infected region.}
\label{fig:FbetaRSIR}
\end{figure}

As an example, Fig.~\ref{fig:FbetaRSIR} shows how $\left\langle s \right\rangle_{gl}(t)$ and $\left\langle i_{tot} \right\rangle_{gl}(t)$ depend on the choice of the homogeneous $\beta$ parameter. This global recovery rate is competing with the heterogeneous infection strengths as defined by the elements of $w$. Increasing its value between the minimal and the maximal matrix element decreases the efficiency of the infection.

\subsubsection{Dependence on the initial conditions}
\begin{figure}[!h]
 \centering  
   \includegraphics[trim=0.5in 1.9in 0.2in 2.5in, width=3.5in]{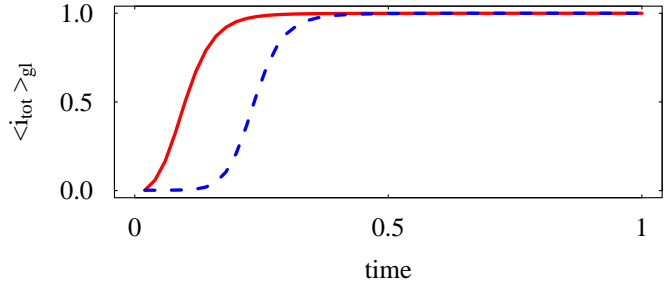}
   \caption{Effect of different source regions in single run simulations on $\left\langle i_{tot} \right\rangle_{gl}(t)$. $\beta=0$ and $F$ were used with initial fully infected region of California (red) and of Monaco (dashed, blue). Both regions belong to the core of $F$.}
   \label{fig:FItotglMonacoVsCA}
\end{figure}

With the $\beta$ parameter being fixed, setting different regions as the source of infection for a single run simulation can lead to different results. The global infection potential of the region depends on its communication environment and its overall influence. As an example, Fig.~\ref{fig:FItotglMonacoVsCA} shows the difference between the global infection curves of two single runs using different sources but otherwise identical settings.

\begin{figure}[!h]
 \centering  
   \includegraphics[trim=0.5in 1.5in 0.2in 2.25in, width=3.4in]{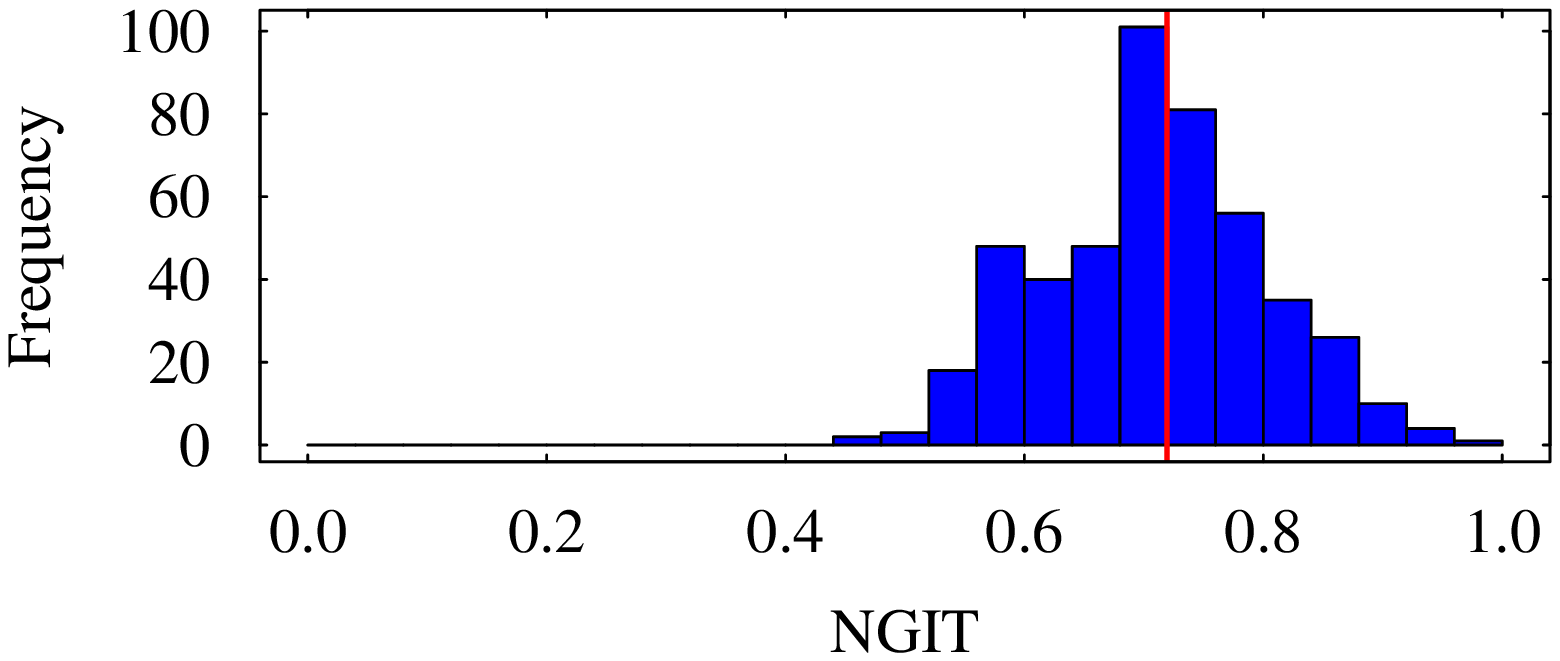}
   \includegraphics[trim=0.5in 1.5in 0.2in 2.25in, width=3.4in]{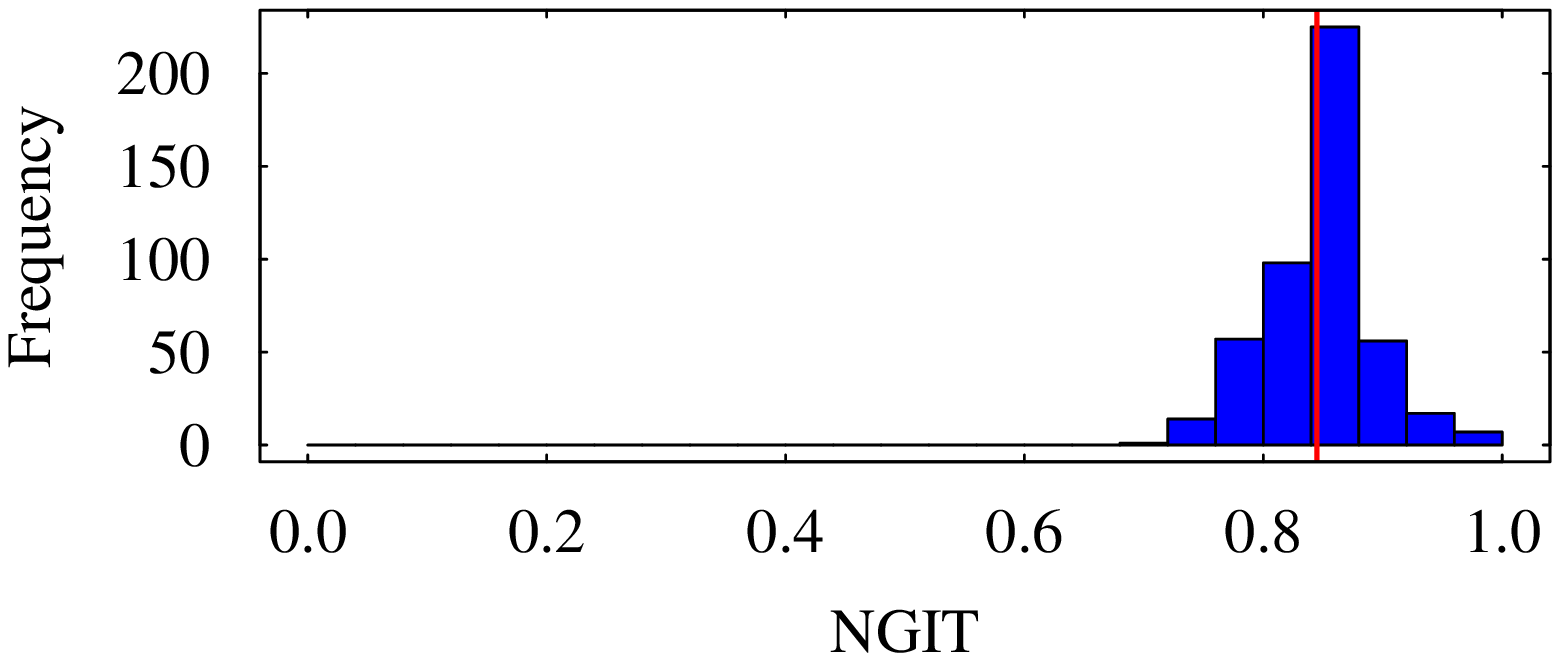}
   \caption{Histograms of the Normalized GIT of single source regions for $F$ (top) and $M$ (bottom). The median values are marked with straight red lines.}
   \label{fig:InvasionHists}
\end{figure}

We ran two series of single run simulations on the $\mathcal{F}$ and the $\mathcal{M}$ networks respectively with each of their regions set as the source.\footnote{The region of \emph{Andaman and Nicobar} was excluded from $\mathcal{M}$ as outlier with low degrees.} The resulting $\mathrm{GIT}_k$ values were then assigned to the regions. Figure~\ref{fig:InvasionHists} shows the distributions of the normalized values obtained after division by the respective maximal $\mathrm{GIT}$ values. The median of the distribution of $\mathrm{NGIT}$ using $F$ is $0.72$ and using $M$ is $0.84$.

\begin{figure}[!h]
 \centering  
   \includegraphics[trim=0.5in 1.55in 0.2in 2.5in, width=3.4in]{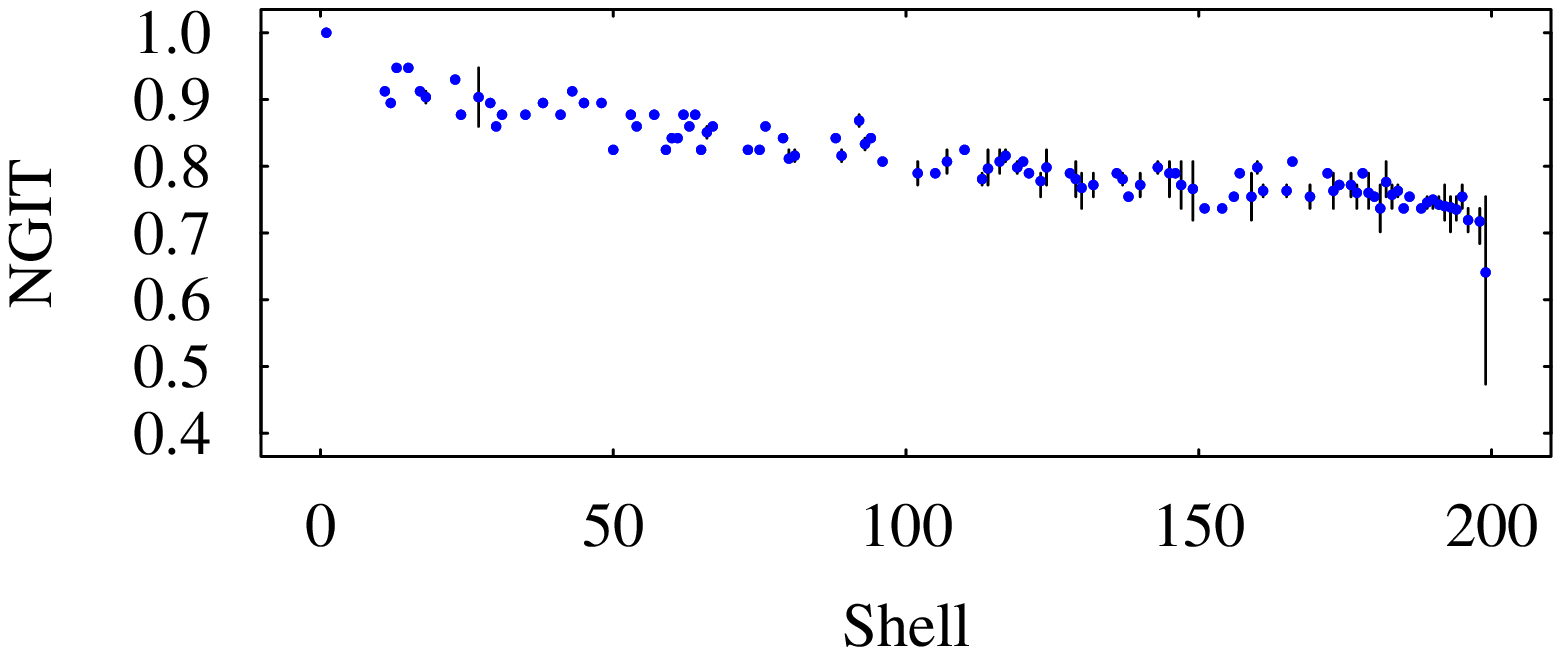}
   \includegraphics[trim=0.5in 1.55in 0.2in 2.5in, width=3.4in]{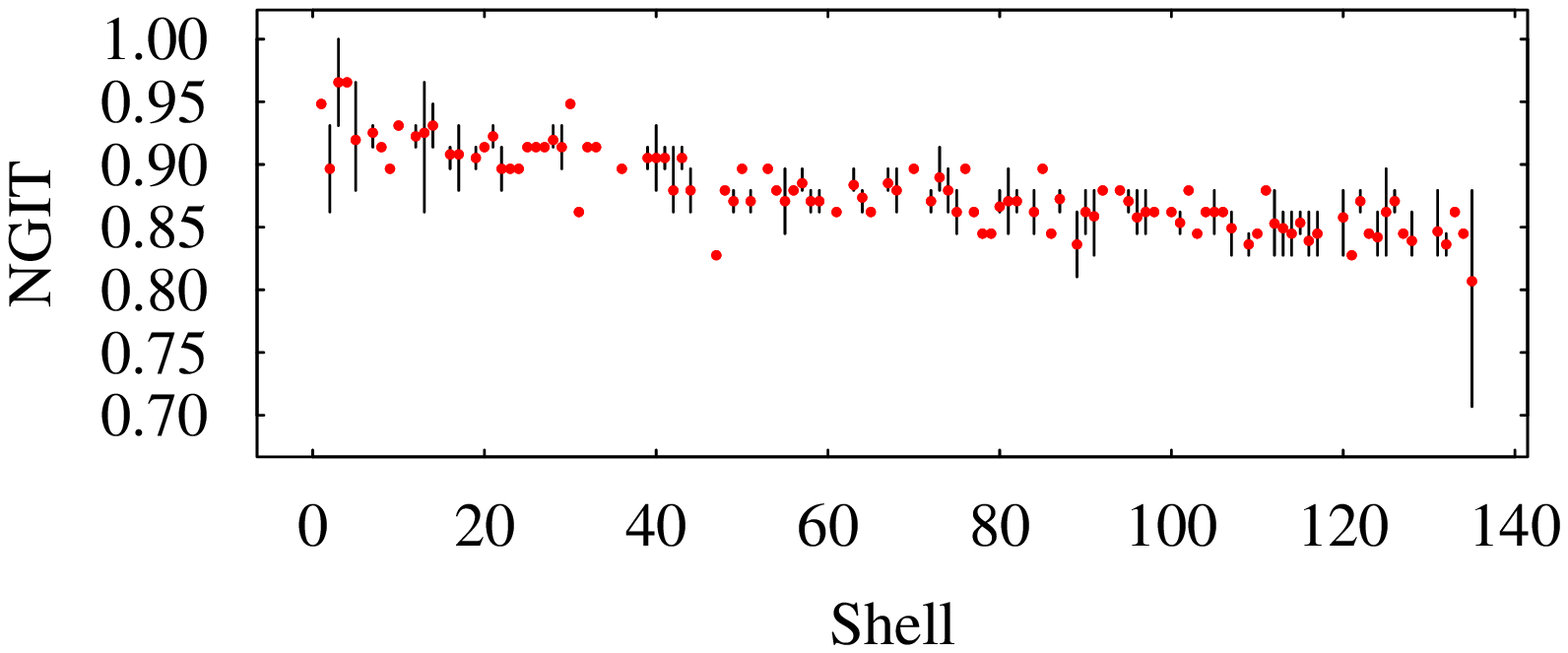}
   \caption{Normalized global infection times of single initial regions for $F$ (top) and $M$ (bottom) as a function of the source region's shell. Dots indicate the mean and a vertical segment connects the minimum and maximum values within each shell.}
   \label{fig:InvasionVSshell}
\end{figure}

We find that the $\mathrm{NGIT}$ value of a region shows a clear negative correlation with its level in the hierarchy determined by the k-shell decomposition. This is illustrated by the Fig.~\ref{fig:InvasionVSshell} for $M$ and $F$ cases. 

\begin{figure}[!h]
 \centering  
   \includegraphics[trim=0in 0in 0in 0in, clip, width=3.5in]{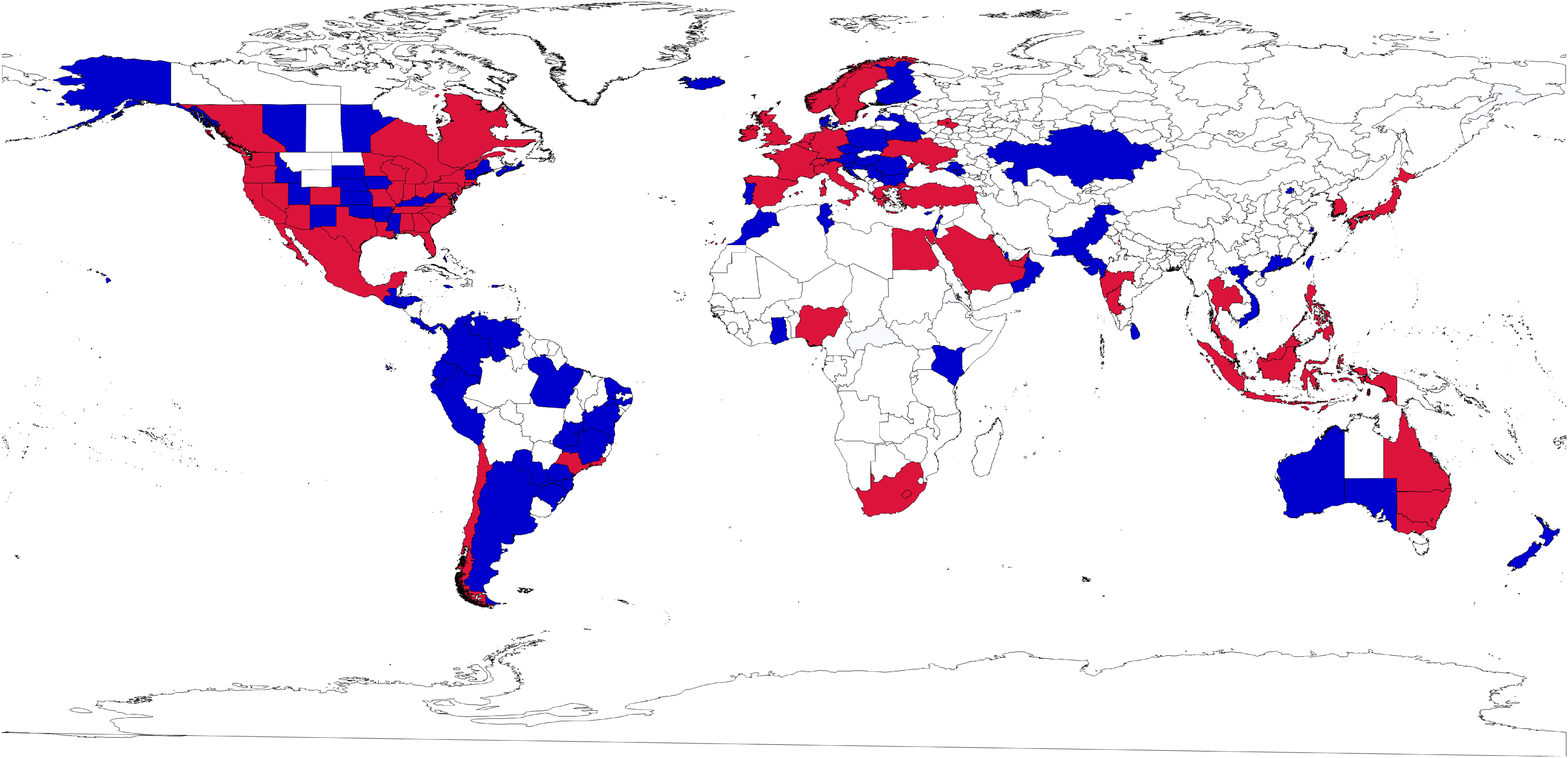}
   \caption{Regions of the Effective Cores. The common regions are shown in red, and the remaining part of $\mathrm{EC}_F$ is blue.}
   \label{fig:EffectiveCoresMap}
\end{figure}

This leads to our last definition. We call the Effective Core ($\mathrm{EC}$) of a regional network the ensemble of regions belonging to the Core and  having $\mathrm{NGIT}$ smaller than any of the periphery values. This limit is $0.68$ for $\mathrm{EC}_F$ and results in a total number of $159$ member regions, while for $\mathrm{EC}_M$ the limit is $0.81$ and there are $72$ regions in it.As shown on Fig.~\ref{fig:EffectiveCoresMap}, we found that the $\mathrm{EC}_M$ is entirely included in the $\mathrm{EC}_F$. 

\section{Conclusion and Future work}
In this paper we analyzed an extensive collection of Twitter conversations between geo-users of the online platform. We defined two aggregated regional communication networks: the Follower and the Mention networks. We analyzed asymmetry and size dependence of
the weighted and directed graph representations, and revealed properties of the regional communication heterogeneity. The collapsed graph representations
were used to determine the Core-Periphery structure
by means of k-shell decomposition. A modified $Regional-SIR$ model was defined and used to further differentiate the regions hierarchy and create the Effective Core of the regions with the greatest global information propagation potentials.

In the future this novel aggregation method can be ameliorated by better data sets. The Core-Periphery division could be determined by a modified decomposition using the original graphs instead of their collapsed counterparts. The $R$-$SIR$ model could be modified to incorporate external influence, that could lead to simulations for heterogeneous parameter fitting to measured real-world processes. This would also require creation measured time series of real infection processes.

\section*{Acknowledgment}
The authors would like to thank the partial support of the European Union and the European Social Fund through project FuturICT.hu (grant no.: TAMOP-4.2.2.C-11/1/KONV-2012-0013), the OTKA 7779 and the NAP 2005/KCKHA005 grants.   EITKIC\_12-1-2012-0001 project was partially supported by the Hungarian Government, managed by the National Development Agency, and financed by the Research and   Technology Innovation Fund and the MAKOG Foundation.



\begin{thebibliography}{1}
\bibitem{Twitter1}
H.~Kwak, C.~Lee, H.~Park, S.~Moon, 
"What is Twitter, a social network or a news media?,"
In \emph{19th ACM WWW}, Raleigh, NC, 2010, pp. 591--600.

\bibitem{Twitter2}
J.~Kulshrestha, F.~Kooti, A.~Nikravesh, and K.P.~Gummadi,
"Geographic dissection of the Twitter network,"
in \emph{6th AAAI ICWSM}, Dublin, 2012.

\bibitem{TwitterDB}
L.~Dobos et.~al.,
"A Multi-terabyte database for geotagged social network data," 
Submitted to the \emph{CogInfoCom 2013}.

\bibitem{kS1}
M.~Kitsak et al., 
"Identification of influential spreaders in complex networks,"
Nat.~Phys. Vol.~\textbf{6}, 888-893 (2010).

\bibitem{kS2}
S.~Carmi, S.~Havlin, S.~Kirkpatrick, Y.~Shavitt, and E.~Shir,
"A model Internet of topology using k-shell decomposition," 
PNAS 104 (27) 1115011154 (2007). 

\bibitem{kS3}
S.~N.~Dorogovtsev, A.~V.~Goltsev, and J.~F.~F.~Mendes,
"k-Core organization of complex networks," 
Phys.~Rev.~Lett. Vol.\textbf{96} 040601 (2006).

\bibitem{sir1}
R.M. Anderson and R.M. May,
"Infectious Diseases of Humans: Dynamics and Control,"
Oxford: Oxford Science Publications, 1992.

\bibitem{sir2}
L. Hufnagel, D. Brockmann, and T. Geisel, 
"Forecast and control of epidemics in a globalized world," 
PNAS 101 (42) 15124-15129 (2004). 

\bibitem{sir3}
C. Castellano, S. Fortunato, and V. Loreto, 
"Statistical physics of social dynamics," 
Rev. Mod. Phys., 81., 591-646 (2009).

\bibitem{2005Densification}
J.~Leskovec, J.~Kleinberg, and C.~Faloutsos,
"Graphs over time: densification laws, shrinking 
diameters and possible explanations,"
in \emph{11th ACM SIGKDD}, Chicago, IL, 2005, pp.177-187.
\end{thebibliography}
%

\end{document}